\documentclass[twocolumn]{aastex63}
\usepackage{graphicx}
\usepackage{amsmath}
\usepackage{amssymb}
\usepackage{cases}
\usepackage{mathrsfs}
\usepackage[flushleft]{threeparttable}
\usepackage{amssymb}
\usepackage{pifont}
\usepackage{amsmath}
\usepackage{epstopdf}
\usepackage{xcolor}
\usepackage[all]{hypcap}
\usepackage{mwe,tikz}
\usepackage{bm}
\usepackage{tabularx}
\usepackage{multirow}
\usepackage{lineno}
\usepackage{xspace}
\usepackage{rotating}

\newcommand{\e}{\text{e}}

\newcommand{\p}{\partial}

\definecolor{red}{RGB}{255, 0, 0}

\submitjournal{ApJL}

\shorttitle{}
\shortauthors{}

\begin{document}

\title{Synchrotron Radiation Dominates the Extremely Bright GRB 221009A}

\correspondingauthor{Bin-Bin Zhang, Xiaohong Zhao, Shao-Lin Xiong}
\email{bbzhang@nju.edu.cn,zhaoxh@ynao.ac.cn,xiongsl@ihep.ac.cn}

\author[0000-0002-5485-5042]{Jun Yang}
\affiliation{School of Astronomy and Space Science, Nanjing University, Nanjing 210093, China}
\affiliation{Key Laboratory of Modern Astronomy and Astrophysics (Nanjing University), Ministry of Education, China}

\author[0000-0003-3659-4800]{Xiao-Hong Zhao}
\affiliation{Yunnan Observatories, Chinese Academy of Sciences, Kunming, China}
\affiliation{Center for Astronomical Mega-Science, Chinese Academy of Sciences, Beijing, China}
\affiliation{Key Laboratory for the Structure and Evolution of Celestial Objects, Chinese Academy of Sciences, Kunming, China}

\author[0009-0008-2841-3065]{Zhenyu Yan}
\affiliation{School of Astronomy and Space Science, Nanjing University, Nanjing 210093, China}
\affiliation{Key Laboratory of Modern Astronomy and Astrophysics (Nanjing University), Ministry of Education, China}

\author[0000-0002-9738-1238]{Xiangyu Ivy Wang}
\affiliation{School of Astronomy and Space Science, Nanjing University, Nanjing 210093, China}
\affiliation{Key Laboratory of Modern Astronomy and Astrophysics (Nanjing University), Ministry of Education, China}

\author{Yan-Qiu Zhang}
\affiliation{Key Laboratory of Particle Astrophysics, Institute of High Energy Physics, Chinese Academy of Sciences, 19B Yuquan Road, Beijing 100049, People’s Republic of China}
\affiliation{University of Chinese Academy of Sciences, Beijing 100049, China}

\author{Zheng-Hua An}
\affiliation{Key Laboratory of Particle Astrophysics, Institute of High Energy Physics, Chinese Academy of Sciences, 19B Yuquan Road, Beijing 100049, People’s Republic of China}

\author{Ce Cai}
\affiliation{College of Physics and Hebei Key Laboratory of Photophysics Research and Application, Hebei Normal University, Shijiazhuang, Hebei 050024, China}

\author{Xin-Qiao Li}
\affiliation{Key Laboratory of Particle Astrophysics, Institute of High Energy Physics, Chinese Academy of Sciences, 19B Yuquan Road, Beijing 100049, People’s Republic of China}

\author{Zihan Li}
\affiliation{School of Astronomy and Space Science, Nanjing University, Nanjing 210093, China}

\author{Jia-Cong Liu}
\affiliation{Key Laboratory of Particle Astrophysics, Institute of High Energy Physics, Chinese Academy of Sciences, 19B Yuquan Road, Beijing 100049, People’s Republic of China}
\affiliation{University of Chinese Academy of Sciences, Beijing 100049, China}

\author{Zi-Ke Liu}
\affiliation{School of Astronomy and Space Science, Nanjing University, Nanjing 210093, China}
\affiliation{Key Laboratory of Modern Astronomy and Astrophysics (Nanjing University), Ministry of Education, China}

\author{Xiang Ma}
\affiliation{Key Laboratory of Particle Astrophysics, Institute of High Energy Physics, Chinese Academy of Sciences, 19B Yuquan Road, Beijing 100049, People’s Republic of China}

\author{Yan-Zhi Meng}
\affiliation{School of Astronomy and Space Science, Nanjing University, Nanjing 210093, China}
\affiliation{Key Laboratory of Modern Astronomy and Astrophysics (Nanjing University), Ministry of Education, China}

\author{Wen-Xi Peng}
\affiliation{Key Laboratory of Particle Astrophysics, Institute of High Energy Physics, Chinese Academy of Sciences, 19B Yuquan Road, Beijing 100049, People’s Republic of China}

\author{Rui Qiao}
\affiliation{Key Laboratory of Particle Astrophysics, Institute of High Energy Physics, Chinese Academy of Sciences, 19B Yuquan Road, Beijing 100049, People’s Republic of China}

\author{Lang Shao}
\affiliation{College of Physics, Hebei Normal University, Shijiazhuang 050024, China}

\author{Li-Ming Song}
\affiliation{Key Laboratory of Particle Astrophysics, Institute of High Energy Physics, Chinese Academy of Sciences, 19B Yuquan Road, Beijing 100049, People’s Republic of China}

\author{Wen-Jun Tan}
\affiliation{Key Laboratory of Particle Astrophysics, Institute of High Energy Physics, Chinese Academy of Sciences, 19B Yuquan Road, Beijing 100049, People’s Republic of China}
\affiliation{University of Chinese Academy of Sciences, Beijing 100049, China}

\author{Ping Wang}
\affiliation{Key Laboratory of Particle Astrophysics, Institute of High Energy Physics, Chinese Academy of Sciences, 19B Yuquan Road, Beijing 100049, People’s Republic of China}

\author{Chen-Wei Wang}
\affiliation{Key Laboratory of Particle Astrophysics, Institute of High Energy Physics, Chinese Academy of Sciences, 19B Yuquan Road, Beijing 100049, People’s Republic of China}
\affiliation{University of Chinese Academy of Sciences, Beijing 100049, China}

\author{Xiang-Yang Wen}
\affiliation{Key Laboratory of Particle Astrophysics, Institute of High Energy Physics, Chinese Academy of Sciences, 19B Yuquan Road, Beijing 100049, People’s Republic of China}

\author{Shuo Xiao}
\affiliation{Guizhou Provincial Key Laboratory of Radio Astronomy and Data Processing, Guizhou Normal University, Guiyang 550001, People’s Republic of China}

\author{Wang-Chen Xue}
\affiliation{Key Laboratory of Particle Astrophysics, Institute of High Energy Physics, Chinese Academy of Sciences, 19B Yuquan Road, Beijing 100049, People’s Republic of China}
\affiliation{University of Chinese Academy of Sciences, Beijing 100049, China}

\author[0000-0003-0691-6688]{Yu-Han Yang}
\affiliation{Department of Physics, University of Rome ``Tor Vergata'', via della Ricerca Scientifica 1, I-00133 Rome, Italy}

\author[0000-0002-5596-5059]{Yihan Yin}
\affiliation{School of Physics, Nanjing University, Nanjing 210093, China}

\author[0000-0002-9725-2524]{Bing Zhang}
\affiliation{Department of Physics and Astronomy, University of Nevada Las Vegas, NV 89154, USA}

\author{Fan Zhang}
\affiliation{Key Laboratory of Particle Astrophysics, Institute of High Energy Physics, Chinese Academy of Sciences, 19B Yuquan Road, Beijing 100049, People’s Republic of China}

\author{Shuai Zhang}
\affiliation{College of Physics, Hebei Normal University, Shijiazhuang 050024, China}

\author{Shuang-Nan Zhang}
\affiliation{Key Laboratory of Particle Astrophysics, Institute of High Energy Physics, Chinese Academy of Sciences, 19B Yuquan Road, Beijing 100049, People’s Republic of China}

\author{Chao Zheng}
\affiliation{Key Laboratory of Particle Astrophysics, Institute of High Energy Physics, Chinese Academy of Sciences, 19B Yuquan Road, Beijing 100049, People’s Republic of China}
\affiliation{University of Chinese Academy of Sciences, Beijing 100049, China}

\author{Shi-Jie Zheng}
\affiliation{Key Laboratory of Particle Astrophysics, Institute of High Energy Physics, Chinese Academy of Sciences, 19B Yuquan Road, Beijing 100049, People’s Republic of China}

% LAST NAME

\author{Shao-Lin Xiong}
\affiliation{Key Laboratory of Particle Astrophysics, Institute of High Energy Physics, Chinese Academy of Sciences, 19B Yuquan Road, Beijing 100049, People’s Republic of China}

\author[0000-0003-4111-5958]{Bin-Bin Zhang}
\affiliation{School of Astronomy and Space Science, Nanjing University, Nanjing 210093, China}
\affiliation{Key Laboratory of Modern Astronomy and Astrophysics (Nanjing University), Ministry of Education, China}
\affiliation{Purple Mountain Observatory, Chinese Academy of Sciences, Nanjing 210023, China}

\begin{abstract}

The brightest Gamma-ray burst, GRB 221009A, has spurred numerous theoretical investigations, with particular attention paid to the origins of ultra-high energy TeV photons during the prompt phase. However, analyzing the mechanism of radiation of photons in the $\sim$MeV range has been difficult because the high flux causes pile-up and saturation effects in most GRB detectors. In this letter, we present systematic modeling of the time-resolved spectra of the GRB using unsaturated data obtained from {\it Fermi}/GBM (precursor) and {\it SATech-01}/GECAM-C (main emission and flare). Our approach incorporates the synchrotron radiation model, which assumes an expanding emission region with relativistic speed and a global magnetic field that decays with radius, and successfully fits such a model to the observational data. Our results indicate that the spectra of the burst are fully in accordance with a synchrotron origin from relativistic electrons accelerated at a large emission radius. The lack of thermal emission in the prompt emission spectra supports a Poynting-flux-dominated jet composition.

\end{abstract}


\section{Introduction} \label{sec:intro}

Despite extensive research spanning several decades, the radiation mechanism of gamma-ray bursts (GRBs) in the prompt phase still remains elusive (see \citealt{Kumar15,Zhang18} for reviews). A typical GRB spectrum can be empirically described as a broken power-law function, namely, the so-called Band function \citep{Band1993ApJ}. The low and high energy slopes are typically $\alpha\sim-1$ and $\beta\sim -2.2$ \citep{Preece2000ApJS, Kaneko2006ApJS}, respectively. The prevalence of non-thermal spectra in GRBs indicates that photosphere emission \citep{2000ApJ...530..292M, 2005ApJ...628..847R, 2010MNRAS.407.1033B, 2010ApJ...725.1137L} is unlikely to be the dominant mechanism. Instead, synchrotron radiation \citep{1998MNRAS.296..275D, 2000MNRAS.313L...1G, 2011A&A...526A.110D, 2014ApJ...784...17B,2020NatAs...4..174B,Zhang20,Wang2022ApJ} appears to be the most favorable explanation for most GRB spectra. The synchrotron origin of prompt emission is also supported by broadband data spanning a wide range of wavelengths, from gamma-rays down to the optical band \citep{2017ApJ...846..137O, Oganesyan2018A&A, Oganesyan2019A&A, Ravasio2018A&A, Ravasio2019A&A}.

However, the measured low energy slope of $\alpha\sim -1$ contradicts the simplest synchrotron model, which assumes a constant magnetic field. \cite{2014NatPh..10..351U} argued that if the GRB emission comes from electrons emitting in a large radius from the central engine, as is the case for models invoking magnetic dissipation in a Poynting-flux-dominated jet \cite[e.g.,][]{zhangyan2011ApJ}, the magnetic field strength would decay as a function of time as the emitter moves to larger distances. Such a model can account for a typical Band spectrum and interpret the GRB data well, as has been shown in direct comparisons between the model and GRB data \citep{Zhang2016ApJ, Zhang2018NA}. Nevertheless, it is in general challenging to compare the models with observational data, as it necessitates the use of bright gamma-ray bursts to obtain finely resolved time-dependent spectra. 

GRB 221009A, which was observed recently on October 9th, 2022, at 13:16:59.99 Coordinated Universal Time (hereafter $T_0$), is notable for being the most luminous and energetic gamma-ray burst ever recorded, owing to its exceptional isotropic-equivalent energy output of approximately $10^{55}~{\rm erg}$ \citep[see also ][]{An2023} and its relatively close distance at a redshift of $z=0.151$ \citep{GCN32686, arXiv230207891}. Furthermore, the detection of ultra-high energy TeV photons associated with this event \citep{GCN32677} has sparked intense debate regarding their origin, encompassing discussions on whether they arise from internal dissipation or external shock and whether they originate from the leptonic or hadronic process \citep{2022arXiv221010673R, 2022arXiv221105754Z, 2022arXiv221012855A, Sato2022arXiv221209266, 2023ApJ...944L..34R, 2023arXiv230211111W}.

GRB 221009A triggered several high-energy missions, including the Gamma-ray Burst Monitor \citep[GBM;][]{meegan2009apj} onboard {\it The Fermi Gamma-Ray Space Telescope} \citep{GCN32636} and GECAM-C onboard {\it The SATech-01 Satellite} \citep{GCN32751}. However, its extraordinary brightness led to some irreparable effects on the data of most detectors, such as data saturation and pulse pile-up. Nevertheless, we were able to accurately capture the full temporal profile and obtain high time-resolution spectra by combining data from {\it Fermi}/GBM and {\it SATech-01}/GECAM-C during the prompt emission.

\begin{figure*}
\label{fig:fig1}
\centering
\includegraphics[width=0.7\textwidth]{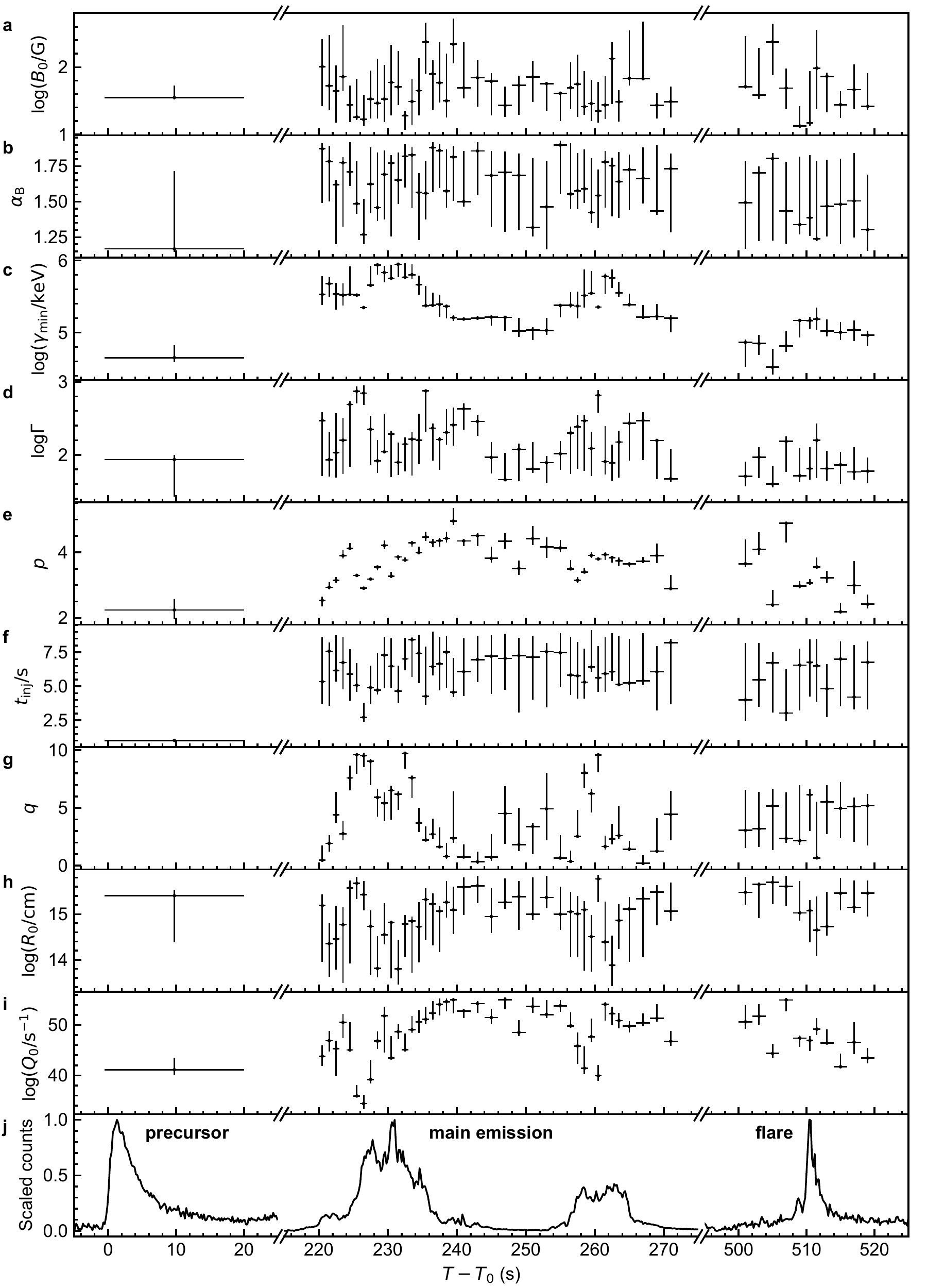}
\caption{The observed light curve of GRB 221009A and its spectral evolution as reflected by the best-fit parameters of our synchrotron model. {\bf a-i}, The best-fit values and 1$\sigma$ uncertainties of the nine model parameters for all time slices in the precursor, main emission, and flare episodes. {\bf j}, The scaled light curves derived from {\it Fermi}/GBM data for precursor, and {\it SATech-01}/GECAM-C data for main emission and flare.}
\end{figure*}

Figure \ref{fig:fig1} demonstrates that the prompt emission phase of GRB 221009A lasts around 600 seconds after T$_{\rm 0}$ and can be segmented into three distinct episodes. The first episode is considered a precursor emission of the burst, which exhibits a fast-rising exponential-decay (FRED) shape and lasts for about 30 seconds. After a quiet period of 180 seconds, the main emission episode appears from 220 to 270 seconds and features two consecutive pulses dominating its temporal profile. Finally, the flare episode takes over, with the majority of its emission concentrated between 500 and 520 seconds. The exceptional intensity of all three episodes presents a unique opportunity to validate the synchrotron model through the use of time-resolved spectral data. In this {\it Letter}, we first conducted a thorough analysis of the observational data by {\it Fermi}/GBM and {\it SATech-01}/GECAM-C (\S \ref{sec:data_ana}). We then expounded on the physical framework of our model in \S \ref{sec:sync_mo}. The fitting procedures were thoroughly outlined in \S \ref{sec:spec_fit}, and the subsequent results and implications were discussed in \S \ref{sec:res_imp}.

\begin{figure*}
\label{fig:fig2}
\centering
\includegraphics[width=0.7\textwidth]{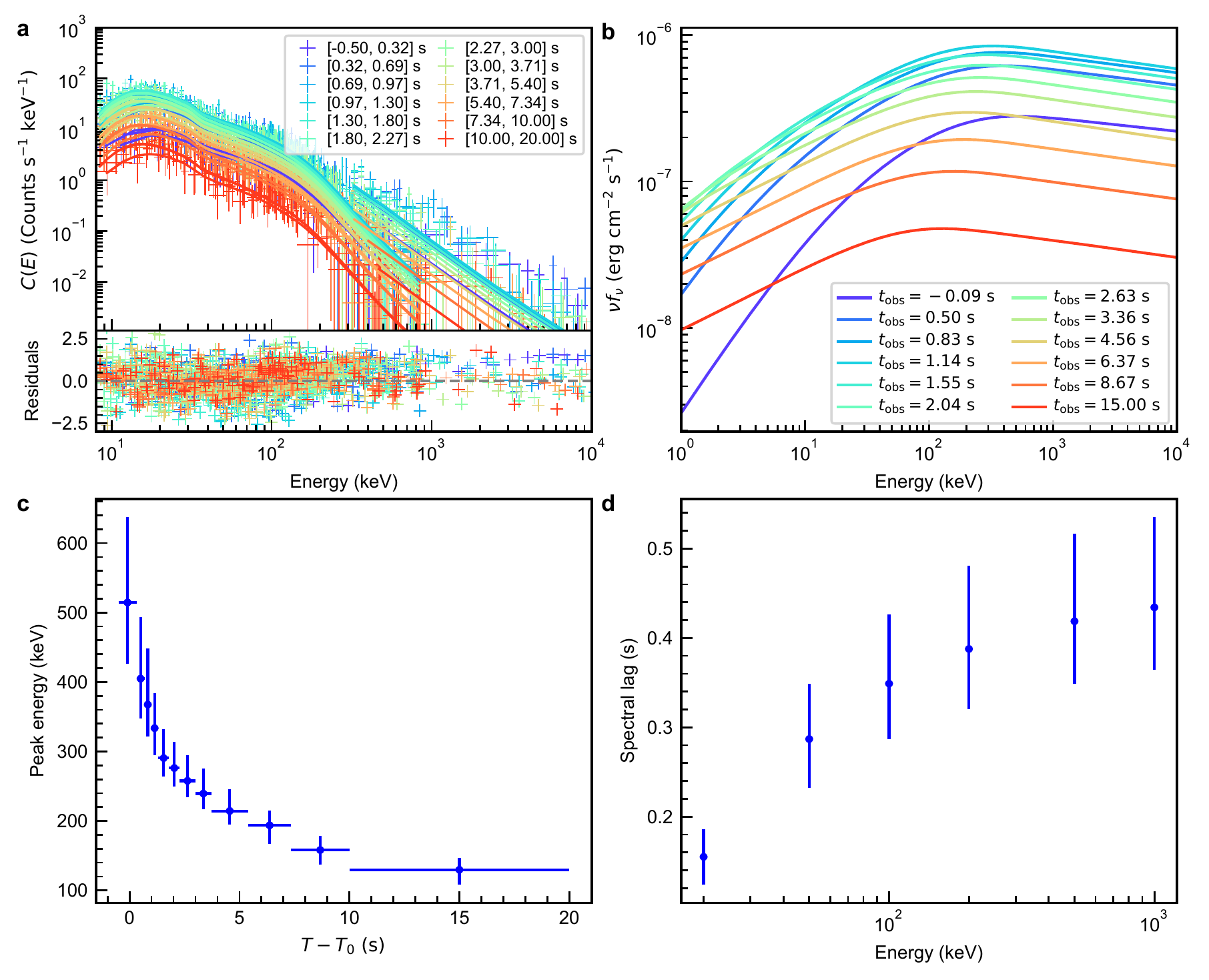}
\caption{The synchrotron fit for the precursor episode. {\bf a}, The observed and modeled photon count spectra. {\bf b}, The evolution of the $\nu f_{\nu}$ spectra as a function of the observed times. {\bf c}, The evolution of peak energies derived from the $\nu f_{\nu}$ spectra. {\bf d}, The evolution of spectral lags as a function of the energies. All error bars represent the 1$\sigma$ confidence level.}
\end{figure*}

\begin{figure*}
\label{fig:fig3}
\centering
\includegraphics[width=0.7\textwidth]{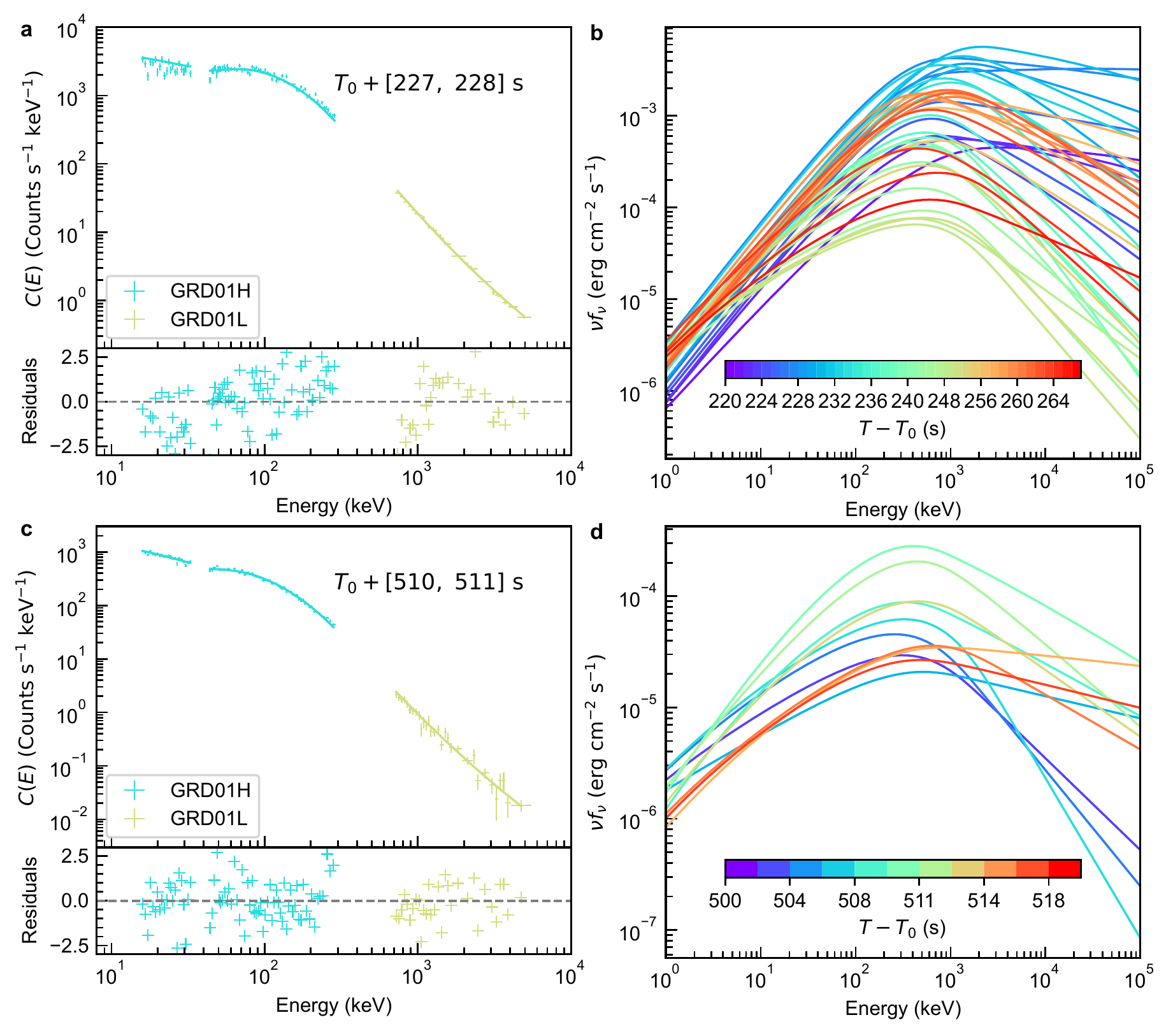}
\caption{The synchrotron fit for the main emission and flare episodes. {\bf a}, The observed and modeled photon count spectra during the brightest time slice of the main emission episode. {\bf b}, The evolution of the $\nu f_{\nu}$ spectra during main emission as a function of the observed times. {\bf c}, The observed and modeled photon count spectra during the brightest time slice of flare episode. {\bf d}, The evolution of the $\nu f_{\nu}$ spectra during the flare as a function of the observed times.}
\end{figure*}

\section{Data Reduction and Analysis} \label{sec:data_ana}

Analyzing the prompt emission of GRB 221009A has been demonstrated to be a challenging task due to its exceptional brightness posing significant electronic disturbances for the majority of GRB detectors, inducing but not limited to data saturation and pulse pile-up effects \citep[e.g.,][]{arxiv230213383}. Fortunately, the moderately bright precursor episode can be accurately recorded by sensitive gamma-ray detectors, such as the {\it Fermi}/GBM, without suffering from the above-mentioned effects \citep{GCN32642}. During the main emission and flare episodes, the GRD01 detector of the {\it SATech-01}/GECAM-C, thanks to its specialized design and special working mode, was confirmed to be capable of avoiding data saturation and pulse pile-up issues, and hence recording precise light curves and spectral shapes \citep{GCN32751}. Therefore, we combine the {\it Fermi}/GBM data from the precursor episode with the {\it SATech-01}/GECAM-C data from the main emission and flare episodes and attempt to explain the complete prompt emission of GRB 221009A using the synchrotron radiation model.

The procedure for data reduction and analysis of {\it Fermi}/GBM data for GRB 221009A followed the same process as described in \cite{Zhang2011ApJ} and \cite{Yang2022Natur}. First, we retrieved the time-tagged event data set covering the time range of GRB 221009A from the {\it Fermi}/GBM public data archive\footnote{\url{https://heasarc.gsfc.nasa.gov/FTP/fermi/data/gbm/daily/}}. Next, we selected three sodium iodide (NaI) detectors (namely n6, n7, and n8) and one bismuth germanium oxide (BGO) detector (b1) with optimal viewing angles for spectral analysis and divided the precursor episode into 12 time slices with equal signal-to-noise levels. For each combination of detector and time slice, the source spectrum and background spectrum were obtained by summing up the number of total photons and background photons for each energy channel, respectively. The number of background photons was determined by simulating the background level using the baseline algorithm\footnote{\url{https://github.com/derb12/pybaselines}} on each energy channel. Furthermore, the detector response matrix in the direction of GRB 221009A was generated using the {\it gbm\_drm\_gen}\footnote{\url{https://github.com/grburgess/gbm_drm_gen}} package \citep{Burgess2018MNRAS, Berlato2019ApJ}.

\cite{An2023} provides a comprehensive description of the data reduction and analysis procedure for {\it SATech-01}/GECAM-C. Here, we highlight some key notes. Each GRD detector contains two independent modes, namely high-gain (6-300 keV) and low-gain (0.4-6 MeV), which cover a considerable energy range spanning three orders of magnitude. During the main emission episode, we partitioned the time range from $T_0+220$ to $T_0+272$ seconds into 40 time slices. For the flare episode, we selected 11 time slices within a 20-second time window around the peak, which contains most of the significant radiation. We then acquired source spectra, background spectra, and response matrices for both high- and low-gain channels for each time slice for spectral analysis. It is worth noting that there is an issue with inaccurate dead time recording in the high-gain data, but this does not distort the spectral shape. Therefore, we utilized the low-gain spectrum as a reference and applied a scaling factor to the high-gain spectrum in each time slice.

\section{The Synchrotron Model} \label{sec:sync_mo}

Consider an ultra-relativistic thin shell ejected from the GRB central engine, within which the magnetic field is entrained with the ejected material and electrons are accelerated into a power-law distribution with an index of $p$, given by $\frac{dN_{\rm e}}{d\gamma_{\rm e}}\propto \gamma_{\rm e}^{-p}$, through mechanisms such as magnetic dissipation \citep{zhangyan2011ApJ}. Upon injection, the electrons will primarily undergo cooling through synchrotron and adiabatic processes towards lower energies, while inverse Compton cooling is typically negligible due to the Klein-Nishina effect and, therefore, not taken into account.
The continuity equation of electrons is 
\begin{equation}
\frac{\partial}{\partial t^{\prime}} \left(\frac{dN_{\rm e}}{d\gamma_{\rm e}} \right) + \frac{\partial}{\partial \gamma_{\rm e}} \left[ \Dot{\gamma}_{\rm e} \left(\frac{dN_{\rm e}}{d\gamma_{\rm e}} \right) \right]=Q\left( \gamma_{\rm e}, t^{\prime} \right),
\end{equation}
where $Q\left( \gamma_{\rm e},t^{\prime}\right)$ is the injection rate of electrons, which is the function of electron energy $\gamma_{\rm e}$ and time $t^{\prime}$ in the co-moving frame of the shell, and reads as
\begin{equation}
Q\left( \gamma_{\rm e},t^{\prime}\right) = 
\left\{\begin{matrix} 
Q_0 { (\frac{t^{\prime}}{{t^\prime_0}} ) }^q \gamma_{\rm e}^{-p}, \quad \gamma_{\rm min} < \gamma_{\rm e} < \gamma_{\rm max} \\
0, \quad \quad \mathrm{otherwise}
\end{matrix}\right. ,
\end{equation}
where $Q_0$ is the injection coefficient, $\gamma_{\rm{min}}$ and $\gamma_{\rm{max}}$ are the minimum and maximum Lorentz factor of the injected electrons, respectively. Here we consider the injection rate increases with a power-law in time \citep{Zhang2016ApJ} and ceases at an observed time $t_{\rm inj}=(1+z)(R_{\rm inj}-R_0)/2\Gamma^2c$, where $R_0$ and $R_{\rm inj}$ are respectively the initial radius where GRB emission begins to be generated and the radius where the injection ceases. Note that we adopt the convention that the co-moving frame, the electron energy, magnetic field ($B$), injection rate, and electron distribution are unprimed although they are measured in the shell co-moving frame. The synchrotron cooling and adiabatic cooling rate are given by
\begin{equation}
\Dot{\gamma}_{\rm e,tot} = \Dot{\gamma}_{\rm e,syn} + \Dot{\gamma}_{\rm e,adi} = -\frac{\sigma_{\rm T} {B}^2 {\gamma_{\rm e}}^2}{6 \pi m_{\rm e} c} - \frac{2\gamma_{\rm e}}{3(t^\prime + t^\prime_0)},
\end{equation}
where $t'_0 =\frac{t_0}{\Gamma}= \frac{R_0}{\Gamma\beta_\Gamma c}$ is the initial time in the co-moving frame of the shell, $t_0$ is the initial time in the burst source frame where GRB emission begins to be generated. In addition, $\Gamma=\frac{1}{\sqrt{1-\beta_\Gamma^2}}$ is the bulk Lorentz factor, c is the light speed, $m_{\rm e}$ is the electron mass, and $\sigma_{\rm T}$ is Thomson scattering cross section. Due to the conservation of magnetic flux, the magnetic field within the shell will decrease as the emission region expands \citep{2001A&A...369..694S}. The exact decay form of the magnetic field depends on the unknown magnetic field configuration. For simplicity, we adopt the following generalized form to describe the magnetic field decay behaviors:
\begin{equation}
 B=B_0(\frac{t'}{t'_0})^{-\alpha_{\rm B}},
\end{equation}
where $B_{\rm 0}$ is the initial magnetic field and $\alpha_{\rm B}$ is the decaying index. 

The synchrotron radiation power in the co-moving frame is \citep{Rybicki&Lightman1979}
\begin{equation}
P^{\prime}\left(\nu^{\prime}\right)=\frac{\sqrt{3} q_{\mathrm{e}}^{3} B}{m_{\mathrm{e}} c^{2}} \int_{\gamma_{\rm min}}^{\gamma_{\rm max}}\left(\frac{d N_{\mathrm{e}}}{d \gamma_{\rm e}}\right) F\left(\frac{\nu^{\prime}}{\nu_{\mathrm{c}}^{\prime}}\right) d \gamma_{\rm e},
\end{equation}
where $\nu_{\mathrm{c}}^{\prime}=3 q_{\mathrm{e}} B \gamma_e^2 /\left(4 \pi m_{\mathrm{e}} c\right)$, $F(x) = x \int_{x}^{+\infty} K_{5/3}(k) dk$, $K_{5/3}(k)$ is the Bessel function, and $q_e$ is the electron charge. Considering the equal-arrival-time surface effect \citep[e.g.,][]{Sari1998}, the observed specific flux can be obtained by
\begin{equation}
F_{\nu_{\mathrm{obs}}}=\frac{1+z}{4 \pi D_{\rm L}^{2}} \int_{t_0}^{t_{\rm e}} 
\frac{c}{2R}
\frac{P^{\prime}\left(\nu^{\prime}\left(\nu_{\mathrm{obs}}\right)\right)}{{\Gamma}^3 {\left( 1-\beta_\Gamma cos\theta\right)}^2} d t,
\end{equation}
where $\nu_{\rm obs} = \nu^{\prime} \mathcal{D}/(1+z)$ is observed photon frequency, $\mathcal{D}=1/[\Gamma (1-\beta_\Gamma cos\theta)]$ is Doppler factor, $\theta$ is the angle between the velocity of an infinitesimal volume of the jet and line of sight, $t_{\rm e} = t_0 + t_{\rm }/(1+z)/(1-\beta _\Gamma)$ is the time in the burst source frame corresponding to observed time $t_{\rm }$ and $D_{\rm L}$ is the luminosity distance obtained by adopting a flat $\Lambda \mathrm{CDM}$ universe with $H_0 =67.4~\mathrm{km}~\mathrm{s}^{-1}~\mathrm{Mpc}^{-1}$, $\Omega_{\rm m} = 0.315$, $\Omega_{\Lambda} = 0.685$ \citep{Plank2020}.

Substituting Eqs. (1-5) to (6), we can obtain the final observed flux predicted by our model in the form of
\begin{eqnarray}
{F_{\nu_{\mathrm{obs}}}} = \nonumber \\ 
 F_{\nu_{\mathrm{obs}}}(t_{\rm },\nu, B_0, \alpha_{\rm B}, \gamma_{\rm min}, \Gamma, p, t_{\rm inj}, q, R_0, Q_0, \gamma_{\rm max}, z ), \label{eq:MBF}
\end{eqnarray}

In this study, we keep $\gamma_{\rm max}$ fixed at $10^{8}$. So final free parameter set, $\mathcal{P}$, of Eq. (7) includes the following 9 terms:
\begin{itemize}
 \item The initial radius $R_0$ in unit of centimeter where the GRB emission begins to be generated.
 \item The initial magnetic field strength $B_0$ in unit of Gauss at the initial radius of the emission region.
 \item The power-law decay index $\alpha_{\rm B}$ of the magnetic field. 
 \item The bulk Lorentz factor $\Gamma$ of the emission region.
 \item The minimum Lorentz factor $\gamma_{\rm min}$ of injected electrons.
 \item The power-law index $p$ of the injected electron spectrum.
 \item The power-law index $q$ of the injection rate of electrons as a function of $t'$. 
 \item The injection time $t_{\rm inj}$ in observer's frame in unit of second.
 \item The electron injection rate coefficient $Q_0$ in units of $\rm s^{-1}$.
\end{itemize}

Adapting the prior bounds as listed in Table \ref{tab:prior}, We can then fit our synchrotron model, namely,
\begin{equation}
F_{\nu_{\mathrm{obs}}}=F_{\nu_{\mathrm{obs}}}(t_{\rm },\nu, \mathcal{P}),
\end{equation}
to the observed spectra at $t$ in the observer frame.

\begin{table}
\centering
\caption{The prior bound of each model parameter for spectral fitting.}
\label{tab:prior}
\begin{tabular}{ccc}
\hline
\hline
\multirow{2}{*}{Parameters} & \multicolumn{2}{c}{Prior bounds} \\
\cline{2-3}
& precursor & main emission \& flare \\
\hline
log$(B_0/{\rm G})$ & [0, 3] & [1, 3] \\
$\alpha_{\rm B}$ & [1, 2] & [1, 2] \\
log$\gamma_{\rm min}$ & [4, 7] & [3, 6] \\
log$\Gamma$ & [1.2, 3.0] & [1.5, 3.0] \\
$p$ & [1.5, 3.5] & [2, 6] \\
$t_{\rm inj}/{\rm s}$ & [-0.5, 2.0] & [0, 10] \\
$q$ & (0) & [0, 10] \\
log$(R_0/{\rm cm})$ & [12, 16] & [12, 16] \\
log$(Q_0/{\rm s^{-1}})$ & [31, 56] & [28, 56] \\
\hline
\hline
\end{tabular}
\end{table}

\section{The Fit} \label{sec:spec_fit}

In accordance with the methodology outlined in \cite{Yang2022Natur}, we utilize our self-developed Python package, {\it MySpecFit}, to fit all the spectral data with our model as described in Eq. (8). {\it MySpecFit} implements Bayesian parameter estimation by wrapping {\it PyMultinest} \citep{Buchner2014A&A}, a Python interface to the popular Fortran nested sampling implementation {\it Multinest} \citep{Feroz2008MNRAS, Feroz2009MNRAS, Buchner2014A&A, Feroz2019OJAp}. {\it Multinest} begins by drawing a set of points in parameter space, called live points, and creating ellipsoids around them. The likelihood is evaluated at each live point, and the point with the lowest likelihood is removed, while new point with higher likelihood is generated in the ellipsoids around the remaining live points, until the exploration ends in a sufficiently small sampling volume. {\it Multinest} excels in sampling and evidence evaluation from distributions that may contain multiple modes and highly degeneracy, and performs well in low to moderate dimensional parameter spaces. In {\it MySpecFit}, PGSTAT\footnote{\url{https://heasarc.gsfc.nasa.gov/xanadu/xspec/}} \citep{Arnaud1996ASPC} is employed as a statistical metric to evaluate the likelihood, which is appropriate for Poisson data in the source spectrum with Gaussian background in the background spectrum.

\subsection{Time-dependent Fit to the Precursor}

To apply Eq. (8) in its simplest form, one can assume that only one single electron ejection event occurs and solve Eq. (8) for each time step to obtain a series of spectra for any given observation time, $t_{\rm obs}$. This approach is only suitable for observation data that has a temporal shape resembling a single pulse, which is the case for the precursor of GRB 221009A. We are thus motivated to fit the observed time-resolved spectra of the complete time series of precursor episode with the time-involved model $F(t_{\rm}, \nu, \mathcal{P})$, where $\mathcal{P}$ represents the single set of parameters described above. 

Interestingly, our initial attempts using broad prior ranges (see Table \ref{tab:prior}) show that the posterior distribution of $q$ was centered around zero, indicating that electrons are injected into the emission region at a constant rate during the precursor episode. Thus, we fix $q$ at zero for the time-resolved fit. After achieving a successful fit with statistically acceptable goodness of fit values (i.e., PGSTAT/d.o.f $\sim$ 1), we listed the best-fit parameters, their 1$\sigma$ uncertainties (see also Figure \ref{fig:fig1}), and fit goodness in Table \ref{tab:specfit}. Figure \ref{fig:fig2}a exhibits the comparison between the data and the model. The observed time-dependent $\nu f_{\nu}$ spectra predicted by the best-fit synchrotron model are displayed in Figures \ref{fig:fig2}b. Furthermore, we reproduce the observed hard-to-soft spectral evolution and hundreds of milliseconds of spectral lags, as shown in Figures \ref{fig:fig2}c and \ref{fig:fig2}d, respectively. Figure \ref{fig:figA1} displays the corresponding corner plot of the posterior probability distributions of the parameters for the fit of the synchrotron model to the precursor. We note that most of the parameters are well constrained, except that $\Gamma$ and $R_0$ exhibit a bimodal distribution. The best-fit values for both $\Gamma$ and $R_0$ fall on the component with a higher probability.

\subsection{Time-independent Fit to Main Emission and Flare}

As depicted in Figure \ref{fig:fig1}, the main emission and flare episodes exhibit intricate and variable temporal profiles that consist of multiple simple pulses superimposed on each other, which implies multiple continuous activities of the central engine. Therefore, it is unrealistic to describe their complete evolutionary features using one set of parameters with a single electron ejection event. Hence, we assume that each time slice corresponds to a completely independent ejection and radiation process and fit them independently using Eq. (8), a method also employed in \cite{Zhang2016ApJ}. This approach enables us to explore the temporal evolution of the model parameters in a slice-wise manner. 

By leaving $q$ free and utilizing the prior bounds listed in Table \ref{tab:prior}, we obtained the best-fit parameter sets, their uncertainties (see also Figure \ref{fig:fig1}), and corresponding statistics, as listed in Table \ref{tab:specfit}. The PGSTAT/d.o.f. values are generally around 1, indicating good fits. Figure \ref{fig:fig1} illustrates the evolution of each best-fit parameter. As examples, in Figures \ref{fig:fig3}a and \ref{fig:fig3}c, we present the observed versus modeled photon count spectra for the brightest time slices during the main emission and flare episodes, respectively. The evolution of the $\nu f_{\nu}$ spectra during the main emission and the flaring episodes as a function of the observed times are shown in Figures \ref{fig:fig3}b and \ref{fig:fig3}d, respectively.

\begin{figure}
\label{fig:fig4}
\centering
\includegraphics[width=0.48\textwidth]{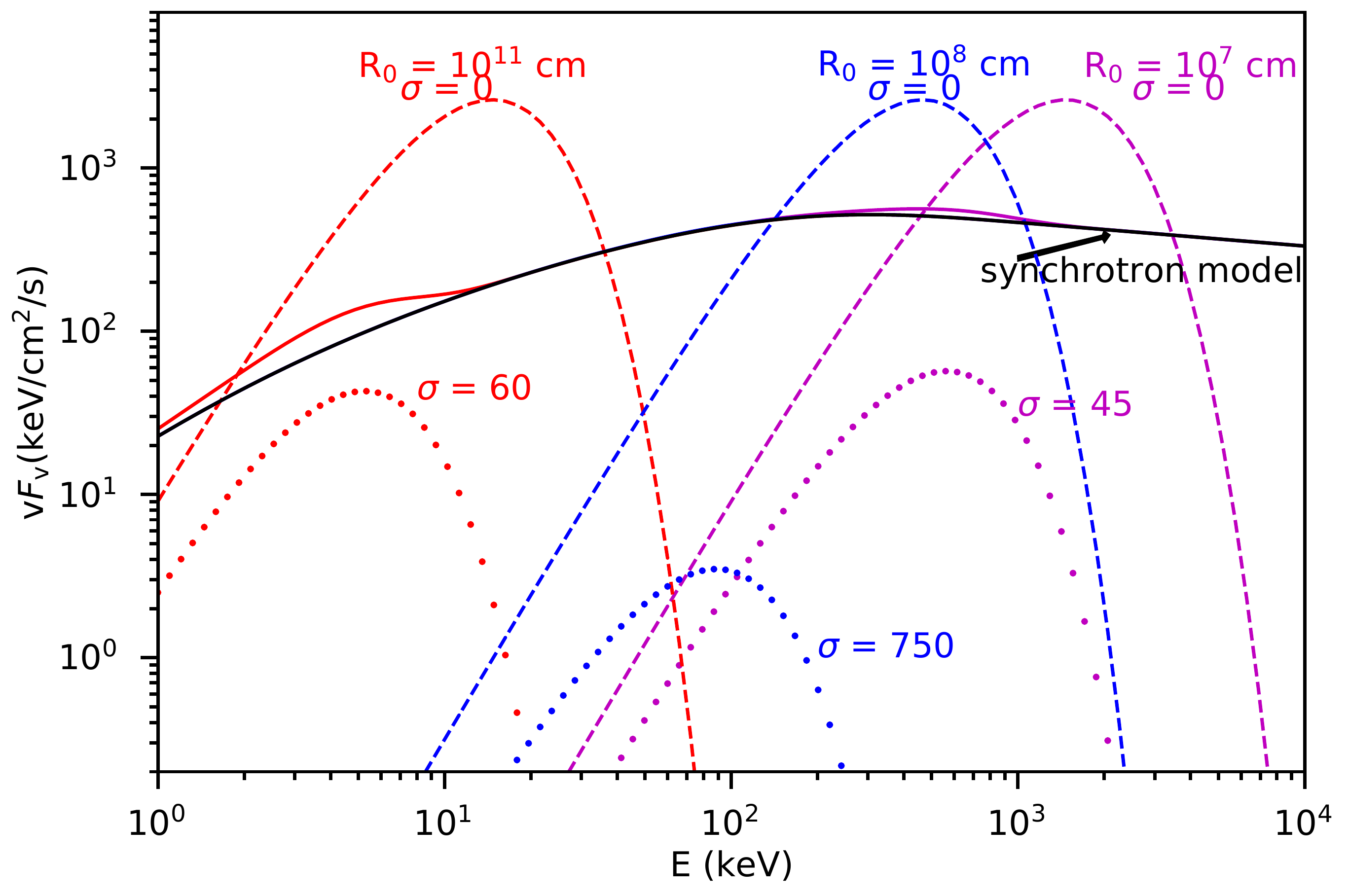}
\caption{Calculation of $\sigma$. The synchrotron spectrum for the time slice between 0.975 and 1.305 s is shown as a black solid line. The colored dashed and dotted lines represent synthetic blackbody components with $\sigma$ values being zero or a lower limit value above which the photosphere emission is not observable, respectively. The colored solid lines represent the hybrid model composed of the synchrotron (black line) and blackbody components (colored dotted line) with the lower limit value of $\sigma$. The unknown size of the jet base at the central engine is adopted as three different values, i.e. $R_{0} = 10^{7},~10^{8},~10^{11}$ cm, which are denoted with the purple, blue and red, respectively. The derived lower limit values of $\sigma$ are $45,~750,~60$, respectively.}
\end{figure}

\begin{figure}
\label{fig:fig5}
\centering
\includegraphics[width=0.48\textwidth]{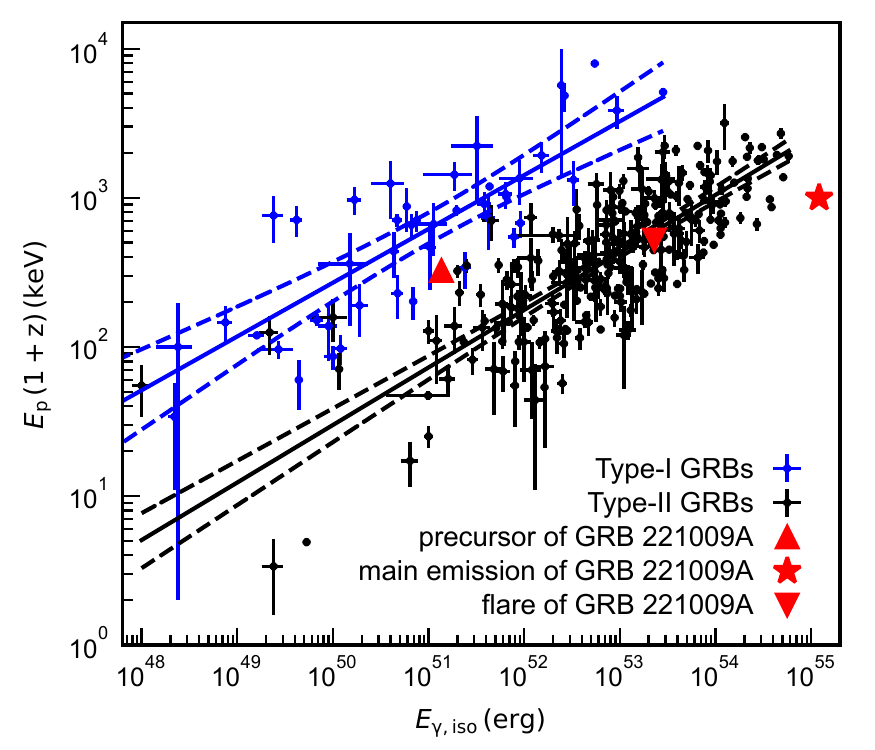}
\caption{The $E_{\rm p}(1+z)$ and $E_{\rm\gamma,iso}$ correlation diagram. The best-fit correlations (solid lines) and corresponding 3$\sigma$ confidence bands (dashed lines) are presented for Type-I (compact star merger origin) and Type-II (massive star core-collapse origin) GRB populations with blue and black colors, respectively (see \citealt{Zhang2009ApJ} for a detailed discussion of the Type I/II classification scheme). The precursor, main emission, and flare episodes of GRB 221009A are denoted by a filled triangle, star, and filled upside-down triangle, respectively. Error bars on data points represent the 1$\sigma$ confidence level.}
\end{figure}

\section{Conclusions and Implications} \label{sec:res_imp}

We successfully fit the observed time-resolved spectra of GRB 221009A using a physical model that incorporates synchrotron radiation of a bulk of relativistic electrons that are accelerated in a large emission region under a decaying magnetic field. Our model successfully reproduced the non-thermal spectra as observed (Figures \ref{fig:fig2} \& \ref{fig:fig3}). The $E_{\rm p}$ values, or the $\nu F{\nu}$ peak, measured by our physical model, fall within the range of $255$ keV to $3.4$ MeV, which is in line with the values presented in \cite{An2023} and \cite{arxiv230213383}. Using the best-fit parameters, our model can also reproduce the observed hard-to-soft spectral evolution and spectral lags during the precursor (see Figure \ref{fig:fig2}).

Our findings indicate that the emission region is approximately $10^{15}$ cm in size, and the magnetic field ranges from a few tens to a few hundred Gauss. This configuration aligns with the scenario that the ejecta is a Poynting-flux-dominated outflow \citep{zhangyan2011ApJ}. Within this scenario, the timescale corresponding to the curvature effect is defined by the duration of the broad pulses. The rapid variability in the lightcurves is related to the mini-jets due to turbulent reconnection in the emission region \citep[e.g.][]{zhangzhang2014,shaogao2022}. Applying the Bayesian method \citep{Scargle2013ApJ}, we derive the shortest variability timescale of about 0.12 s, which is much shorter than the timescale defined by the emission radius. This is fully consistent with the ICMART picture of \cite{zhangyan2011ApJ}.

The Poynting-flux-dominated nature of the outflow can also be demonstrated by calculating the ratio of the Poynting flux's luminosity to the baryonic flux's luminosity, denoted by $\sigma$ \citep{Zhang&Asaf2009apj}. Specifically, $\sigma$ is defined as $\sigma \equiv L_{\rm P}/L_{\rm b}$. A high value of $\sigma$ indicates that the Poynting flux is the primary energy source. If the flow is baryonic flux dominated, one would observe a blackbody spectrum with a temperature estimated as $T_{\rm ph}^{\rm ob} = (L_{\rm w}/4\pi R_{0}^{2}ca)^{1/4}(1+z)^{-1}$, where $L_{\rm w}$ is the initial wind luminosity of the fireball, $R_{0}$ is the radius of the fireball base, $c$ is the speed of light, and $a$ is the Stefan-Boltzmann energy density constant. We tested this hypothesis by setting the initial wind luminosity $L_{\rm w}$ to be equal to the gamma-ray luminosity $L_{\gamma}$ in the time slice between $T_0 + 0.975$ s and $T_0 + 1.305$ s of the precursor, and calculating the blackbody spectrum at $R_{0}$ = $10^{7}, 10^{8}, 10^{11}$ cm, respectively, which are represented by dashed lines in Figure \ref{fig:fig4}. Obviously, all of these spectra had significant thermal-like peaks that were not present in the observed data.

Next, we investigated the maximal level at which the baryonic flux is allowed so the blackbody component, if any, is barely suppressed. This approach allows placing a lower limit onto $\sigma$ \citep{Zhang&Asaf2009apj}. To do so, we first generate the blackbody spectrum given the different $\sigma$ by replacing $L_{\rm w}$ with $L_{\gamma}$/(1 + $\sigma$), as plotted as dotted lines in Figure \ref{fig:fig4}. We added this blackbody spectrum to our best-fit physical spectrum and determined its goodness of fit to the observed data. By using the Akaike Information Criterion \citep[AIC;][]{Akaike_1974ITAC, Sugiura1978FurtherAO}, we could determine the $\sigma$ value at which the hybrid model deviated significantly from the observation \citep[$\Delta$AIC $>$ 5;][]{Krishak2020jacp}. Our calculations with different $R_0$ values all yielded a global lower limit of $\sigma\ge 45$, which strongly suggests that the outflow is dominated by Poynting flux.

By using the average $E_{\rm p}$ and flux values for each episode, we can determine the burst energies and plot them on the $E_{\rm p,z}$-$E_{\gamma,\rm iso}$ diagram \citep{Amati2002A&A, Zhang2009ApJ, Minaev2020MNRAS}, which is depicted in Figure \ref{fig:fig5}. Notably, even with the energy of only the main emission considered, GRB 221009A ranks as the most energetic burst with $E_{\gamma,\rm iso}\sim 1.21\times10^{55}$ erg \citep[see also ][]{An2023}, despite being an extraordinary GRB that follows the same track as other type-II GRBs in the diagram.

\acknowledgments

We acknowledge the support by the National Key Research and Development Programs of China (2018YFA0404204, 2022YFF0711404, 2022SKA0130102), the National Natural Science Foundation of China (Grant Nos. 11833003, U2038105, U1831135, 12121003), the science research grants from the China Manned Space Project with NO.CMS-CSST-2021-B11, the Program for Innovative Talents, and the Strategic Priority Research Program on Space Science, the Chinese Academy of Sciences, grant No. XDB23040400.

\appendix
\restartappendixnumbering

\section{The posterior probability distributions for the fit to the precursor}

\begin{figure*}
\label{fig:figA1}
\centering
\includegraphics[width=0.7\textwidth]{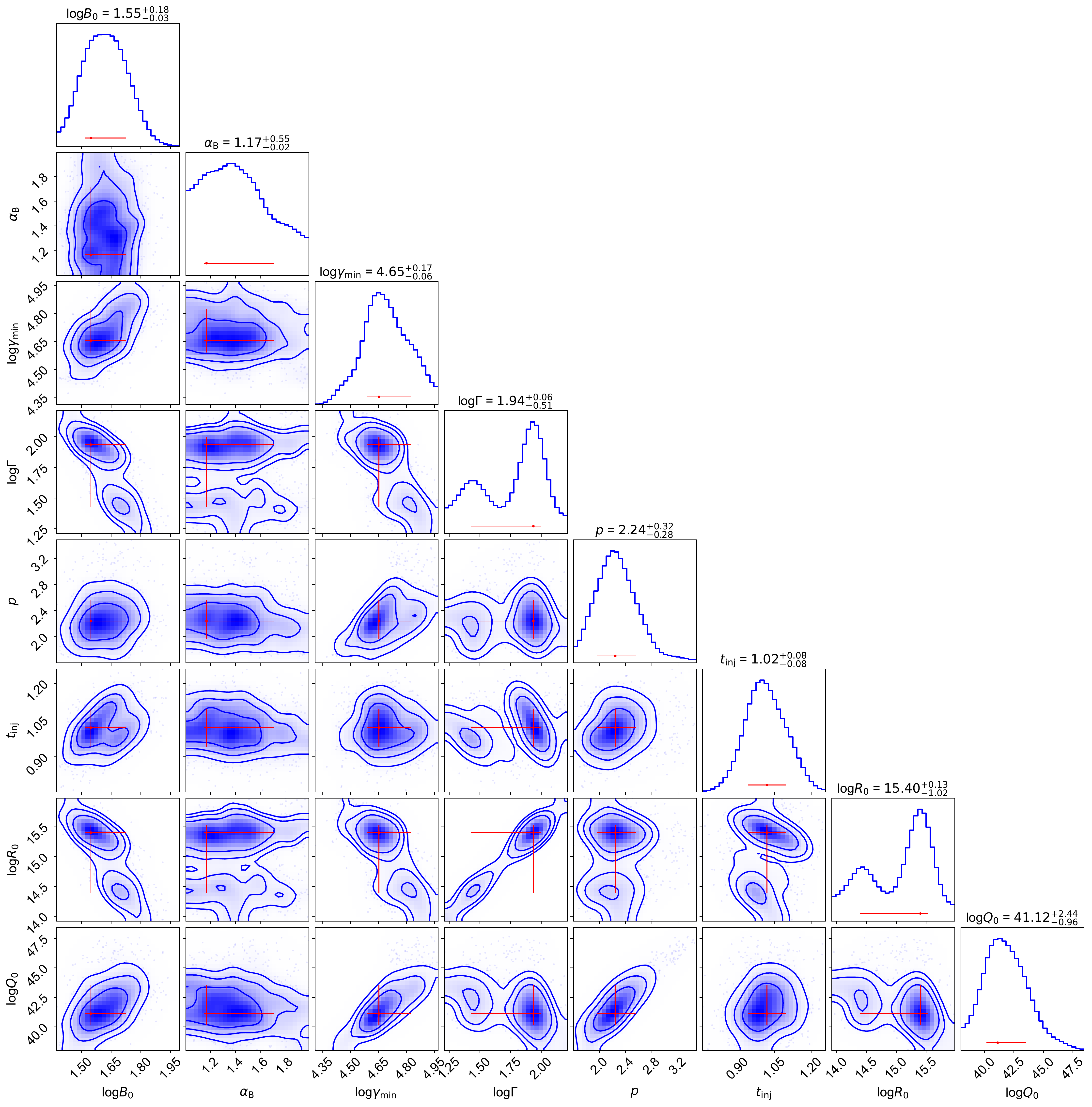}
\caption{Corner plot of the posterior probability distributions of the parameters for the fit of the synchrotron model to the precursor. The red error bars represent 1$\sigma$ uncertainties.}
\end{figure*}

Figure \ref{fig:figA1} displays the corner plot of the posterior probability distributions of the parameters for the fit of the synchrotron model to the precursor episode.

\section{Spectral fitting results}

The best-fit parameters, their 1$\sigma$ uncertainties, and corresponding fit goodness are listed in Table \ref{tab:specfit}. We also list the peak energies in the $\nu f_{\nu}$ spectra derived from our physical models.

\begin{table*}
\tiny
\centering
\setlength\tabcolsep{3pt}
\caption{Spectral fitting results and derived peak energies in each time slice. All errors represent the 1$\sigma$ uncertainties.}
\label{tab:specfit}
\begin{tabular}{rrccccccccrcl}
\hline
\hline
t1 (s) & t2 (s) & log$(B_0/{\rm G})$ & $\alpha_{\rm B}$ & log$\gamma_{\rm min}$ & log$\Gamma$ & $p$ & $t_{\rm inj}/{\rm s}$ & $q$ & log$(R_0/{\rm cm})$ & log$(Q_0/{\rm s^{-1}})$ & PGSTAT/d.o.f. & $E_{\rm p}/{\rm keV}$ \\
\hline
-0.50&20.00&$1.55_{-0.03}^{+0.18}$&$1.17_{-0.02}^{+0.55}$&$4.65_{-0.06}^{+0.17}$&$1.94_{-0.51}^{+0.06}$&$2.24_{-0.28}^{+0.32}$&$1.02_{-0.08}^{+0.08}$&(0)&$15.40_{-1.02}^{+0.13}$&$41.12_{-0.96}^{+2.44}$&4087.90/5824.00&$281.70\pm13.79$ \\
%-0.50&20.00&$1.63_{-0.06}^{+0.13}$&$1.51_{-0.36}^{+0.30}$&$4.73_{-0.07}^{+0.14}$&$1.66_{-0.22}^{+0.07}$&$2.29_{-0.32}^{+0.58}$&$1.01_{-0.06}^{+0.11}$&(0)&$14.85_{-0.44}^{+0.15}$&$42.23_{-1.56}^{+3.42}$&4083.44/5824.00&$260.01\pm11.69$ \\
\hline
220.00&221.00&$2.01_{-0.59}^{+0.41}$&$1.87_{-0.38}^{+0.04}$&$5.53_{-0.15}^{+0.25}$&$2.47_{-0.76}^{+0.11}$&$2.53_{-0.19}^{+0.11}$&$5.34_{-1.61}^{+3.26}$&$0.47_{-0.03}^{+1.26}$&$15.19_{-1.23}^{+0.24}$&$43.80_{-1.89}^{+2.21}$&273.95/128.00&$3409.62_{-547.48}^{+4168.73}$ \\
221.00&222.00&$1.72_{-0.37}^{+0.76}$&$1.78_{-0.28}^{+0.11}$&$5.68_{-0.18}^{+0.09}$&$1.93_{-0.23}^{+0.38}$&$2.93_{-0.05}^{+0.17}$&$7.58_{-4.01}^{+0.63}$&$1.93_{-0.72}^{+0.73}$&$14.35_{-0.73}^{+0.45}$&$46.90_{-2.14}^{+1.89}$&201.86/128.00&$1179.27_{-37.42}^{+110.63}$ \\
222.00&223.00&$1.65_{-0.47}^{+0.38}$&$1.62_{-0.42}^{+0.03}$&$5.54_{-0.22}^{+0.15}$&$2.03_{-0.34}^{+0.54}$&$3.15_{-0.08}^{+0.10}$&$6.16_{-0.80}^{+2.58}$&$4.39_{-0.67}^{+2.01}$&$14.45_{-0.67}^{+0.74}$&$45.31_{-5.33}^{+1.56}$&275.35/128.00&$773.71_{-5.33}^{+72.71}$ \\
223.00&224.00&$1.86_{-0.22}^{+0.77}$&$1.77_{-0.45}^{+0.04}$&$5.52_{-0.15}^{+0.16}$&$2.20_{-0.46}^{+0.30}$&$3.90_{-0.09}^{+0.16}$&$6.74_{-1.96}^{+1.90}$&$2.77_{-0.58}^{+1.14}$&$14.77_{-1.28}^{+0.39}$&$50.51_{-3.10}^{+1.72}$&203.28/128.00&$666.14_{-17.05}^{+23.41}$ \\
224.00&225.00&$1.44_{-0.25}^{+0.29}$&$1.71_{-0.10}^{+0.21}$&$5.53_{-0.01}^{+0.39}$&$2.69_{-0.85}^{+0.04}$&$4.12_{-0.04}^{+0.16}$&$5.90_{-1.96}^{+1.83}$&$7.60_{-1.08}^{+1.09}$&$15.58_{-1.66}^{+0.16}$&$45.05_{-0.25}^{+5.57}$&249.74/128.00&$627.42_{-5.75}^{+17.58}$ \\
225.00&226.00&$1.26_{-0.05}^{+0.57}$&$1.49_{-0.07}^{+0.30}$&$5.52_{-0.01}^{+0.01}$&$2.87_{-0.17}^{+0.06}$&$3.29_{-0.04}^{+0.03}$&$5.08_{-0.49}^{+1.63}$&$9.61_{-1.65}^{+0.01}$&$15.68_{-0.34}^{+0.14}$&$35.92_{-0.24}^{+2.22}$&397.10/128.00&$902.80_{-8.28}^{+23.16}$ \\
226.00&227.00&$1.22_{-0.09}^{+0.37}$&$1.27_{-0.07}^{+0.25}$&$5.35_{-0.01}^{+0.01}$&$2.85_{-0.16}^{+0.10}$&$2.90_{-0.04}^{+0.01}$&$2.72_{-0.31}^{+1.06}$&$9.52_{-0.99}^{+0.27}$&$15.43_{-0.34}^{+0.30}$&$34.40_{-0.99}^{+1.71}$&397.10/128.00&unconstrained \\
227.00&228.00&$1.53_{-0.29}^{+0.42}$&$1.62_{-0.21}^{+0.26}$&$5.66_{-0.00}^{+0.27}$&$2.35_{-0.48}^{+0.19}$&$3.18_{-0.04}^{+0.05}$&$4.91_{-1.22}^{+1.62}$&$9.04_{-2.09}^{+0.19}$&$14.73_{-1.07}^{+0.35}$&$39.18_{-0.80}^{+3.92}$&307.21/128.00&$[1053.43,~1126.19]$ \\
228.00&229.00&$1.47_{-0.23}^{+0.66}$&$1.46_{-0.09}^{+0.39}$&$5.94_{-0.13}^{+0.02}$&$1.92_{-0.16}^{+0.29}$&$3.55_{-0.11}^{+0.01}$&$4.72_{-0.33}^{+2.72}$&$5.91_{-1.68}^{+0.73}$&$13.81_{-0.20}^{+0.70}$&$46.83_{-1.53}^{+1.90}$&258.35/128.00&$[1505.34,~1579.94]$ \\
229.00&230.00&$1.53_{-0.34}^{+0.51}$&$1.69_{-0.32}^{+0.17}$&$5.83_{-0.14}^{+0.09}$&$2.04_{-0.02}^{+0.52}$&$4.21_{-0.11}^{+0.15}$&$7.29_{-2.44}^{+1.35}$&$5.42_{-1.58}^{+0.94}$&$14.55_{-0.21}^{+0.69}$&$51.81_{-7.27}^{+1.86}$&205.38/128.00&$1491.54_{-23.85}^{+38.27}$ \\
230.00&231.00&$1.77_{-0.49}^{+0.56}$&$1.77_{-0.51}^{+0.06}$&$5.75_{-0.03}^{+0.19}$&$2.29_{-0.59}^{+0.04}$&$3.27_{-0.06}^{+0.13}$&$6.48_{-1.58}^{+2.02}$&$6.52_{-2.55}^{+0.42}$&$14.82_{-1.23}^{+0.03}$&$43.46_{-0.27}^{+4.32}$&186.54/128.00&$[2156.09,~2289.14]$ \\
231.00&232.00&$1.71_{-0.27}^{+0.53}$&$1.65_{-0.17}^{+0.19}$&$5.95_{-0.18}^{+0.00}$&$1.90_{-0.17}^{+0.27}$&$3.86_{-0.09}^{+0.03}$&$4.65_{-0.85}^{+1.85}$&$6.19_{-1.34}^{+0.23}$&$13.80_{-0.34}^{+0.64}$&$48.68_{-1.44}^{+1.44}$&253.97/128.00&$1338.52_{-6.15}^{+37.51}$ \\
232.00&233.00&$1.28_{-0.21}^{+0.07}$&$1.82_{-0.22}^{+0.09}$&$5.77_{-0.01}^{+0.19}$&$2.15_{-0.37}^{+0.09}$&$3.76_{-0.01}^{+0.07}$&$7.01_{-0.85}^{+1.55}$&$9.73_{-1.35}^{+0.06}$&$14.78_{-0.73}^{+0.20}$&$45.09_{-0.38}^{+3.22}$&324.62/128.00&$780.87_{-1.80}^{+18.19}$ \\
233.00&234.00&$1.49_{-0.35}^{+0.33}$&$1.83_{-0.38}^{+0.05}$&$5.80_{-0.07}^{+0.14}$&$2.21_{-0.50}^{+0.10}$&$4.29_{-0.11}^{+0.04}$&$8.43_{-2.75}^{+0.19}$&$7.61_{-1.76}^{+0.20}$&$14.85_{-1.13}^{+0.10}$&$49.05_{-0.63}^{+2.77}$&240.13/128.00&$854.25_{-9.78}^{+11.89}$ \\
234.00&235.00&$1.65_{-0.37}^{+0.48}$&$1.56_{-0.34}^{+0.14}$&$5.66_{-0.14}^{+0.13}$&$2.20_{-0.33}^{+0.36}$&$3.99_{-0.04}^{+0.18}$&$7.43_{-2.19}^{+1.34}$&$3.71_{-0.78}^{+1.30}$&$14.72_{-0.78}^{+0.54}$&$50.62_{-2.16}^{+2.28}$&287.75/128.00&$1001.38_{-11.47}^{+23.33}$ \\
235.00&236.00&$2.38_{-0.47}^{+0.28}$&$1.56_{-0.18}^{+0.23}$&$5.37_{-0.00}^{+0.27}$&$2.87_{-0.56}^{+0.00}$&$4.47_{-0.12}^{+0.16}$&$4.27_{-0.63}^{+3.65}$&$2.23_{-0.02}^{+1.19}$&$15.32_{-0.88}^{+0.25}$&$51.14_{-0.98}^{+2.36}$&381.81/128.00&$978.58_{-6.74}^{+29.74}$ \\
236.00&237.00&$1.90_{-0.54}^{+0.20}$&$1.88_{-0.31}^{+0.04}$&$5.38_{-0.01}^{+0.18}$&$2.37_{-0.44}^{+0.06}$&$4.30_{-0.14}^{+0.10}$&$6.45_{-0.65}^{+2.59}$&$2.73_{-0.42}^{+1.35}$&$15.22_{-0.91}^{+0.23}$&$52.37_{-1.25}^{+2.37}$&260.79/128.00&$606.12_{-2.79}^{+21.30}$ \\
237.00&238.00&$1.77_{-0.29}^{+0.60}$&$1.86_{-0.26}^{+0.05}$&$5.39_{-0.17}^{+0.13}$&$2.21_{-0.41}^{+0.03}$&$4.35_{-0.18}^{+0.08}$&$6.66_{-1.95}^{+2.00}$&$1.63_{-0.10}^{+1.68}$&$15.07_{-0.90}^{+0.12}$&$54.09_{-2.17}^{+1.21}$&225.66/128.00&$527.89_{-7.24}^{+20.08}$ \\
238.00&239.00&$1.50_{-0.25}^{+0.70}$&$1.57_{-0.12}^{+0.29}$&$5.37_{-0.17}^{+0.02}$&$2.31_{-0.17}^{+0.32}$&$4.42_{-0.12}^{+0.20}$&$7.51_{-2.62}^{+1.22}$&$0.82_{-0.23}^{+1.16}$&$15.26_{-0.34}^{+0.46}$&$54.63_{-1.94}^{+0.67}$&211.74/128.00&$513.50_{-16.29}^{+11.96}$ \\
239.00&240.00&$2.34_{-0.28}^{+0.38}$&$1.81_{-0.31}^{+0.08}$&$5.21_{-0.05}^{+0.03}$&$2.41_{-0.28}^{+0.24}$&$4.95_{-0.12}^{+0.40}$&$4.56_{-0.36}^{+2.52}$&$2.38_{-0.41}^{+4.04}$&$15.09_{-0.53}^{+0.51}$&$55.00_{-2.69}^{+0.34}$&172.58/128.00&$569.58_{-11.68}^{+18.66}$ \\
240.00&242.00&$1.69_{-0.15}^{+0.68}$&$1.50_{-0.04}^{+0.35}$&$5.19_{-0.00}^{+0.03}$&$2.63_{-0.28}^{+0.07}$&$4.35_{-0.16}^{+0.06}$&$6.07_{-1.78}^{+2.46}$&$0.75_{-0.10}^{+1.08}$&$15.59_{-0.61}^{+0.22}$&$52.80_{-1.39}^{+0.37}$&204.72/128.00&$436.04_{-14.81}^{+4.04}$ \\
242.00&244.00&$1.84_{-0.42}^{+0.27}$&$1.86_{-0.31}^{+0.06}$&$5.20_{-0.01}^{+0.03}$&$2.46_{-0.20}^{+0.09}$&$4.50_{-0.32}^{+0.08}$&$6.96_{-1.64}^{+1.77}$&$0.34_{-0.18}^{+0.89}$&$15.62_{-0.38}^{+0.22}$&$54.26_{-1.81}^{+0.48}$&144.11/128.00&$436.04_{-14.81}^{+4.04}$ \\
244.00&246.00&$1.79_{-0.50}^{+0.12}$&$1.68_{-0.36}^{+0.17}$&$5.21_{-0.11}^{+0.03}$&$1.96_{-0.22}^{+0.21}$&$3.81_{-0.13}^{+0.36}$&$7.21_{-2.78}^{+1.55}$&$0.73_{-0.31}^{+1.98}$&$14.95_{-0.36}^{+0.62}$&$51.48_{-1.29}^{+1.74}$&168.11/128.00&$453.45_{-10.32}^{+25.77}$ \\
246.00&248.00&$1.43_{-0.17}^{+0.28}$&$1.71_{-0.42}^{+0.15}$&$5.21_{-0.19}^{+0.01}$&$1.66_{-0.01}^{+0.36}$&$4.34_{-0.24}^{+0.24}$&$7.05_{-2.32}^{+1.82}$&$4.52_{-2.63}^{+2.38}$&$15.26_{-0.15}^{+0.52}$&$55.03_{-1.91}^{+0.47}$&153.13/128.00&$507.62_{-19.89}^{+26.82}$ \\
248.00&250.00&$1.73_{-0.44}^{+0.14}$&$1.69_{-0.39}^{+0.15}$&$5.02_{-0.08}^{+0.18}$&$2.08_{-0.44}^{+0.07}$&$3.51_{-0.20}^{+0.24}$&$7.25_{-4.16}^{+1.30}$&$1.82_{-0.59}^{+3.20}$&$15.38_{-0.73}^{+0.34}$&$48.53_{-0.76}^{+2.48}$&132.36/128.00&$398.58_{-12.65}^{+33.46}$ \\
250.00&252.00&$1.85_{-0.37}^{+0.24}$&$1.32_{-0.06}^{+0.49}$&$5.04_{-0.14}^{+0.04}$&$1.81_{-0.05}^{+0.37}$&$4.42_{-0.20}^{+0.38}$&$7.15_{-3.11}^{+1.55}$&$3.39_{-2.38}^{+0.31}$&$15.00_{-0.13}^{+0.80}$&$53.67_{-1.28}^{+1.80}$&151.41/128.00&$420.27_{-17.07}^{+29.03}$ \\
252.00&254.00&$1.75_{-0.40}^{+0.03}$&$1.46_{-0.30}^{+0.32}$&$5.03_{-0.06}^{+0.17}$&$1.89_{-0.28}^{+0.09}$&$4.16_{-0.34}^{+0.29}$&$7.54_{-4.09}^{+0.64}$&$4.92_{-2.73}^{+3.13}$&$15.37_{-0.24}^{+0.47}$&$52.09_{-0.70}^{+2.88}$&114.79/128.00&$420.27_{-17.07}^{+29.03}$ \\
254.00&256.00&$1.61_{-0.41}^{+0.01}$&$1.90_{-0.34}^{+0.03}$&$5.38_{-0.16}^{+0.02}$&$2.01_{-0.22}^{+0.18}$&$4.13_{-0.13}^{+0.24}$&$7.46_{-2.48}^{+1.46}$&$0.66_{-0.08}^{+1.98}$&$15.00_{-0.35}^{+0.60}$&$53.82_{-1.19}^{+1.16}$&194.01/128.00&$543.94_{-13.61}^{+16.53}$ \\
256.00&257.00&$1.69_{-0.31}^{+0.37}$&$1.55_{-0.10}^{+0.36}$&$5.38_{-0.03}^{+0.18}$&$2.30_{-0.56}^{+0.09}$&$3.49_{-0.04}^{+0.27}$&$5.82_{-1.46}^{+2.79}$&$0.37_{-0.13}^{+0.93}$&$15.05_{-0.99}^{+0.44}$&$49.83_{-0.33}^{+2.70}$&199.43/128.00&$747.44_{-23.76}^{+22.72}$ \\
257.00&258.00&$1.75_{-0.49}^{+0.44}$&$1.58_{-0.13}^{+0.27}$&$5.37_{-0.17}^{+0.20}$&$2.39_{-0.57}^{+0.16}$&$3.15_{-0.09}^{+0.11}$&$5.77_{-1.66}^{+2.50}$&$2.53_{-0.05}^{+2.31}$&$15.01_{-0.94}^{+0.39}$&$45.85_{-3.58}^{+2.70}$&256.85/128.00&$747.44_{-23.76}^{+22.72}$ \\
258.00&259.00&$1.41_{-0.12}^{+0.53}$&$1.59_{-0.10}^{+0.28}$&$5.51_{-0.17}^{+0.36}$&$2.46_{-0.68}^{+0.08}$&$3.40_{-0.04}^{+0.12}$&$5.31_{-1.22}^{+2.46}$&$8.04_{-1.29}^{+0.80}$&$15.09_{-1.33}^{+0.19}$&$41.43_{-1.19}^{+4.35}$&312.69/128.00&$[683.23,~725.39]$ \\
259.00&260.00&$1.46_{-0.26}^{+0.35}$&$1.42_{-0.07}^{+0.38}$&$5.55_{-0.02}^{+0.31}$&$2.09_{-0.38}^{+0.23}$&$3.90_{-0.08}^{+0.09}$&$6.41_{-0.33}^{+2.71}$&$6.24_{-1.63}^{+0.39}$&$14.51_{-0.77}^{+0.47}$&$47.69_{-1.12}^{+3.56}$&271.69/128.00&$640.56_{-1.47}^{+22.52}$ \\
260.00&261.00&$1.35_{-0.17}^{+0.43}$&$1.54_{-0.23}^{+0.18}$&$5.36_{-0.02}^{+0.00}$&$2.82_{-0.24}^{+0.07}$&$3.80_{-0.04}^{+0.04}$&$5.62_{-1.04}^{+2.34}$&$9.60_{-1.49}^{+0.04}$&$15.77_{-0.51}^{+0.09}$&$39.93_{-1.06}^{+2.18}$&328.50/128.00&$[477.02,~490.39]$ \\
261.00&262.00&$1.44_{-0.21}^{+0.43}$&$1.78_{-0.32}^{+0.08}$&$5.78_{-0.25}^{+0.04}$&$1.91_{-0.18}^{+0.49}$&$3.93_{-0.16}^{+0.08}$&$5.93_{-1.33}^{+2.66}$&$1.66_{-0.25}^{+0.92}$&$14.39_{-0.43}^{+0.89}$&$54.09_{-3.19}^{+0.44}$&299.94/128.00&$1073.02_{-7.39}^{+35.16}$ \\
262.00&263.00&$2.12_{-0.67}^{+0.25}$&$1.75_{-0.36}^{+0.06}$&$5.76_{-0.14}^{+0.12}$&$1.89_{-0.24}^{+0.25}$&$3.84_{-0.16}^{+0.05}$&$6.07_{-0.70}^{+2.85}$&$2.32_{-0.47}^{+1.32}$&$13.88_{-0.46}^{+0.65}$&$52.28_{-2.34}^{+1.36}$&365.89/128.00&$930.24_{-4.27}^{+30.48}$ \\
263.00&264.00&$1.49_{-0.29}^{+0.17}$&$1.64_{-0.25}^{+0.21}$&$5.55_{-0.05}^{+0.16}$&$2.17_{-0.49}^{+0.16}$&$3.74_{-0.15}^{+0.10}$&$5.12_{-0.09}^{+3.58}$&$2.60_{-0.28}^{+2.57}$&$14.86_{-0.63}^{+0.36}$&$50.89_{-1.56}^{+1.51}$&268.71/128.00&$1063.18_{-7.32}^{+37.37}$ \\
264.00&266.00&$1.83_{-0.11}^{+0.71}$&$1.73_{-0.29}^{+0.11}$&$5.39_{-0.02}^{+0.15}$&$2.43_{-0.51}^{+0.14}$&$3.64_{-0.08}^{+0.05}$&$5.24_{-0.61}^{+3.24}$&$1.43_{-0.16}^{+0.72}$&$15.11_{-1.16}^{+0.26}$&$49.77_{-1.25}^{+1.00}$&311.02/128.00&$1063.18_{-7.32}^{+37.37}$ \\
266.00&268.00&$1.83_{-0.02}^{+0.85}$&$1.66_{-0.18}^{+0.22}$&$5.21_{-0.02}^{+0.17}$&$2.47_{-0.54}^{+0.12}$&$3.73_{-0.06}^{+0.13}$&$5.40_{-0.28}^{+3.51}$&$0.20_{-0.03}^{+0.69}$&$15.34_{-1.28}^{+0.25}$&$50.42_{-0.72}^{+1.92}$&177.88/128.00&$459.76_{-3.17}^{+19.46}$ \\
268.00&270.00&$1.44_{-0.20}^{+0.18}$&$1.43_{-0.03}^{+0.46}$&$5.22_{-0.05}^{+0.17}$&$2.20_{-0.53}^{+0.01}$&$3.90_{-0.23}^{+0.36}$&$6.07_{-2.83}^{+1.89}$&$1.26_{-0.19}^{+2.84}$&$15.49_{-0.73}^{+0.15}$&$51.33_{-0.70}^{+2.85}$&220.05/128.00&$733.79_{-13.40}^{+41.70}$ \\
270.00&272.00&$1.49_{-0.23}^{+0.22}$&$1.73_{-0.45}^{+0.11}$&$5.20_{-0.19}^{+0.04}$&$1.67_{-0.04}^{+0.40}$&$2.89_{-0.02}^{+0.41}$&$8.21_{-4.54}^{+0.29}$&$4.44_{-2.25}^{+2.04}$&$15.06_{-0.22}^{+0.63}$&$46.78_{-0.92}^{+2.04}$&147.27/128.00&$[606.12,~675.41]$ \\
\hline
500.00&502.00&$1.71_{-0.03}^{+0.75}$&$1.49_{-0.32}^{+0.29}$&$4.87_{-0.34}^{+0.04}$&$1.71_{-0.14}^{+0.20}$&$3.65_{-0.11}^{+0.75}$&$4.00_{-1.53}^{+4.18}$&$3.06_{-1.55}^{+3.50}$&$15.48_{-0.26}^{+0.35}$&$50.64_{-1.36}^{+3.34}$&174.23/128.00&$307.96_{-3.53}^{+54.71}$ \\
502.00&504.00&$1.58_{-0.06}^{+0.70}$&$1.70_{-0.48}^{+0.05}$&$4.85_{-0.16}^{+0.12}$&$1.97_{-0.29}^{+0.14}$&$4.09_{-0.17}^{+0.52}$&$5.47_{-1.98}^{+2.72}$&$3.19_{-1.60}^{+3.20}$&$15.65_{-0.74}^{+0.04}$&$51.74_{-1.64}^{+2.45}$&142.02/128.00&$255.55_{-20.87}^{+17.02}$ \\
504.00&506.00&$2.38_{-0.50}^{+0.27}$&$1.80_{-0.58}^{+0.04}$&$4.52_{-0.10}^{+0.25}$&$1.60_{-0.04}^{+0.25}$&$2.40_{-0.07}^{+0.45}$&$6.72_{-3.66}^{+0.80}$&$5.17_{-3.81}^{+1.47}$&$15.70_{-0.48}^{+0.14}$&$44.37_{-0.98}^{+1.87}$&158.14/128.00&$505.29_{-53.92}^{+46.52}$ \\
506.00&508.00&$1.69_{-0.32}^{+0.29}$&$1.43_{-0.23}^{+0.35}$&$4.82_{-0.08}^{+0.20}$&$2.18_{-0.42}^{+0.07}$&$4.89_{-0.58}^{+0.02}$&$3.04_{-0.62}^{+3.22}$&$2.35_{-0.37}^{+4.01}$&$15.61_{-0.42}^{+0.23}$&$54.98_{-2.28}^{+0.45}$&155.73/128.00&$323.97_{-29.19}^{+6.02}$ \\
508.00&510.00&$1.13_{-0.03}^{+0.29}$&$1.34_{-0.07}^{+0.48}$&$5.17_{-0.24}^{+0.00}$&$1.71_{-0.09}^{+0.38}$&$2.97_{-0.07}^{+0.16}$&$6.55_{-3.32}^{+1.21}$&$2.16_{-0.38}^{+4.83}$&$15.03_{-0.17}^{+0.70}$&$47.41_{-1.72}^{+0.52}$&169.79/128.00&$339.24_{-26.27}^{+17.63}$ \\
510.00&511.00&$1.17_{-0.04}^{+0.77}$&$1.39_{-0.13}^{+0.44}$&$5.17_{-0.11}^{+0.05}$&$1.81_{-0.15}^{+0.29}$&$3.07_{-0.06}^{+0.11}$&$6.75_{-2.52}^{+1.69}$&$6.13_{-3.13}^{+0.48}$&$15.08_{-0.69}^{+0.23}$&$46.93_{-2.09}^{+0.83}$&158.93/128.00&$407.87_{-33.32}^{+5.68}$ \\
511.00&512.00&$1.99_{-0.61}^{+0.57}$&$1.24_{-0.01}^{+0.55}$&$5.19_{-0.15}^{+0.16}$&$2.20_{-0.52}^{+0.23}$&$3.55_{-0.06}^{+0.29}$&$6.51_{-2.65}^{+1.96}$&$0.67_{-0.02}^{+4.86}$&$14.65_{-0.58}^{+0.74}$&$49.18_{-1.19}^{+2.19}$&147.35/128.00&$443.13_{-15.05}^{+31.70}$ \\
512.00&514.00&$1.86_{-0.54}^{+0.06}$&$1.47_{-0.21}^{+0.33}$&$5.02_{-0.07}^{+0.16}$&$1.81_{-0.17}^{+0.24}$&$3.22_{-0.14}^{+0.22}$&$4.82_{-2.09}^{+1.83}$&$5.53_{-2.82}^{+1.46}$&$14.73_{-0.20}^{+0.76}$&$46.43_{-0.24}^{+2.75}$&131.44/128.00&$447.23_{-27.93}^{+21.08}$ \\
514.00&516.00&$1.44_{-0.20}^{+0.20}$&$1.48_{-0.28}^{+0.32}$&$5.00_{-0.11}^{+0.14}$&$1.86_{-0.26}^{+0.17}$&$2.19_{-0.03}^{+0.28}$&$7.01_{-3.48}^{+1.29}$&$4.97_{-2.61}^{+2.29}$&$15.46_{-0.37}^{+0.33}$&$41.72_{-0.10}^{+2.56}$&180.11/128.00&$848.37_{-129.63}^{+121.24}$ \\
516.00&518.00&$1.67_{-0.28}^{+0.37}$&$1.50_{-0.26}^{+0.34}$&$5.04_{-0.16}^{+0.13}$&$1.77_{-0.10}^{+0.34}$&$2.99_{-0.27}^{+0.74}$&$4.21_{-0.91}^{+3.84}$&$5.13_{-3.07}^{+0.78}$&$15.15_{-0.12}^{+0.59}$&$46.59_{-2.38}^{+4.01}$&140.25/128.00&$643.52_{-28.97}^{+110.84}$ \\
518.00&520.00&$1.42_{-0.05}^{+0.49}$&$1.30_{-0.15}^{+0.39}$&$4.96_{-0.15}^{+0.06}$&$1.78_{-0.17}^{+0.19}$&$2.42_{-0.13}^{+0.29}$&$6.77_{-3.47}^{+1.53}$&$5.19_{-3.45}^{+1.01}$&$15.46_{-0.52}^{+0.28}$&$43.46_{-1.04}^{+2.02}$&168.44/128.00&$481.43_{-13.12}^{+134.53}$ \\
\hline
\hline
\end{tabular}
\end{table*}

\end{document}